\documentclass[summary]{ursi}
\usepackage[numbers]{natbib}
\usepackage{url}

\title{Machine Learning Frameworks for Large-Scale Radio Surveys: A Summary of Recent Studies}

\author{Nikhel Gupta\affref{ref1}}

\affiliation{%
  \aff{ref1}{CSIRO Space \& Astronomy, PO Box 1130, Bentley WA 6102, Australia}
}

\begin{document}

\maketitle

\begin{abstract}
The rapid growth of large-scale radio surveys, generating over 100 petabytes of data annually, has created a pressing need for automated data analysis methods.
Recent research has explored the application of machine learning techniques to address the challenges associated with detecting and classifying radio galaxies, as well as discovering peculiar radio sources.
This paper provides an overview of our investigations with the Evolutionary Map of the Universe (EMU) survey, detailing the methodologies employed—including supervised, unsupervised, self-supervised, and weakly supervised learning approaches--and their implications for ongoing and future radio astronomical surveys.
\end{abstract}

\section{Introduction}

The exploration of the cosmos has been revolutionized by large-scale radio surveys, enabling deeper insights into the Universe. Advanced radio interferometers, such as the Australian Square Kilometre Array Pathfinder (ASKAP; \citep{johnston07ASKAP,DeBoer09,hotan21}), the Murchison Widefield Array (MWA; \citep{tingay13,wayth18}), MeerKAT \citep{jonas16}, the Low Frequency Array (LOFAR; \citep{vanharleem13}), and the Karl G. Jansky Very Large Array (JVLA; \citep{perley11}), play a crucial role in conducting these surveys. Among them, the Evolutionary Map of the Universe (EMU; \citep{hopkins25}) survey, carried out with ASKAP, exemplifies the transformative impact of modern radio interferometers. Over five years, EMU aims to detect about 20 million compact and extended radio galaxies, producing an extensive dataset that will refine our understanding of galaxy evolution and the Universe’s history. Furthermore, this wealth of data is expected to reveal novel astrophysical phenomena and provide deeper insights into the origins of radio emissions. However, to fully harness its potential, traditional data analysis methods must be supplemented with advanced techniques.

In response to the challenges posed by the vast data volumes generated by next-generation radio telescopes, machine learning has emerged as a key analytical tool (e.g., \citep{mostert21, gupta22, walmsley22, segal22, gupta23a, lochner23, gupta23b, slijepcevic23, Mohale24, gupta24a, Lastufka24, gupta24b, riggi24, lochner24, mostert24, lao25, gupta25a, gupta25b}). These methods have significantly accelerated the identification of new radio morphologies while enhancing the efficiency of source detection, classification, and cataloguing. 
This paper explores various machine learning techniques applied to radio astronomical data from the EMU survey, emphasizing the role of labeled datasets. 
We discuss different approaches, including supervised, unsupervised, self-supervised, and weakly-supervised learning, demonstrating their impact on large-scale radio surveys.
This paper is structured as follows: Section~\ref{SEC:emu} describes the EMU survey. Section~\ref{SEC:methods} outlines the machine learning techniques employed, along with a summary of the investigations conducted using the EMU survey data. Section~\ref{SEC:conclusions} gives a brief conclusion.

\section{ASKAP's EMU Survey}
\label{SEC:emu}
The EMU survey aims to develop a comprehensive radio atlas of the southern sky using the advanced ASKAP radio telescope. 
The survey will result in 853 tile footprints from 1,014 individual tile observations. Of these, 692 tiles will have a 10-hour integration time, while 161 tiles, observed twice, will have a 5-hour integration time. 
The 10-hour integration observations will achieve a median RMS noise of $30~\mu$Jy/beam and typically detect around 750 sources per square degree, while the 5-hour integrations will have a median RMS noise of $46~\mu$Jy/beam, detecting about 460 sources per square degree. 
The five-year survey will cover the 800–1088 MHz frequency range centered at 944 MHz. 
EMU will map declinations from $-11.4^{\circ}$ to $+7.0^{\circ}$ and is expected to catalogue approximately 20 million radio sources across $2\pi$ steradians, with completion anticipated by 2028.
In the first year of EMU observations, a total of 160 tiles were observed, covering about 4,500 square degrees of the sky. 
Data collection for the survey commenced in late 2022, with validated data being delivered between February 2023 and March 2024. 
These datasets include tiles with Scheduling Block unique ID (SBID) numbers ranging from 45638 to 59612, and are accessible via the CSIRO Data Access Portal (CASDA\footnote{https://research.csiro.au/casda/}).
Prior to the main EMU survey, the first EMU Pilot Survey (EMU-PS1; \citep{norris21}) was conducted in late 2019, covering an area of 270 deg$^2$ in the region of $301^{\circ} < {\rm RA} < 336^{\circ}$ and $-63^{\circ} < {\rm Dec} < -48^{\circ}$, composed of 10 tiles with a total integration time of about 10 hours per tile.

\section{Machine Learning Methods}
\label{SEC:methods}
In this section, we discuss different machine learning techniques employed to analyse the data from the EMU-PS1 and the first year of the main EMU survey.

\begin{figure}[!htbp]
  \centering
  \includegraphics[width=80mm]{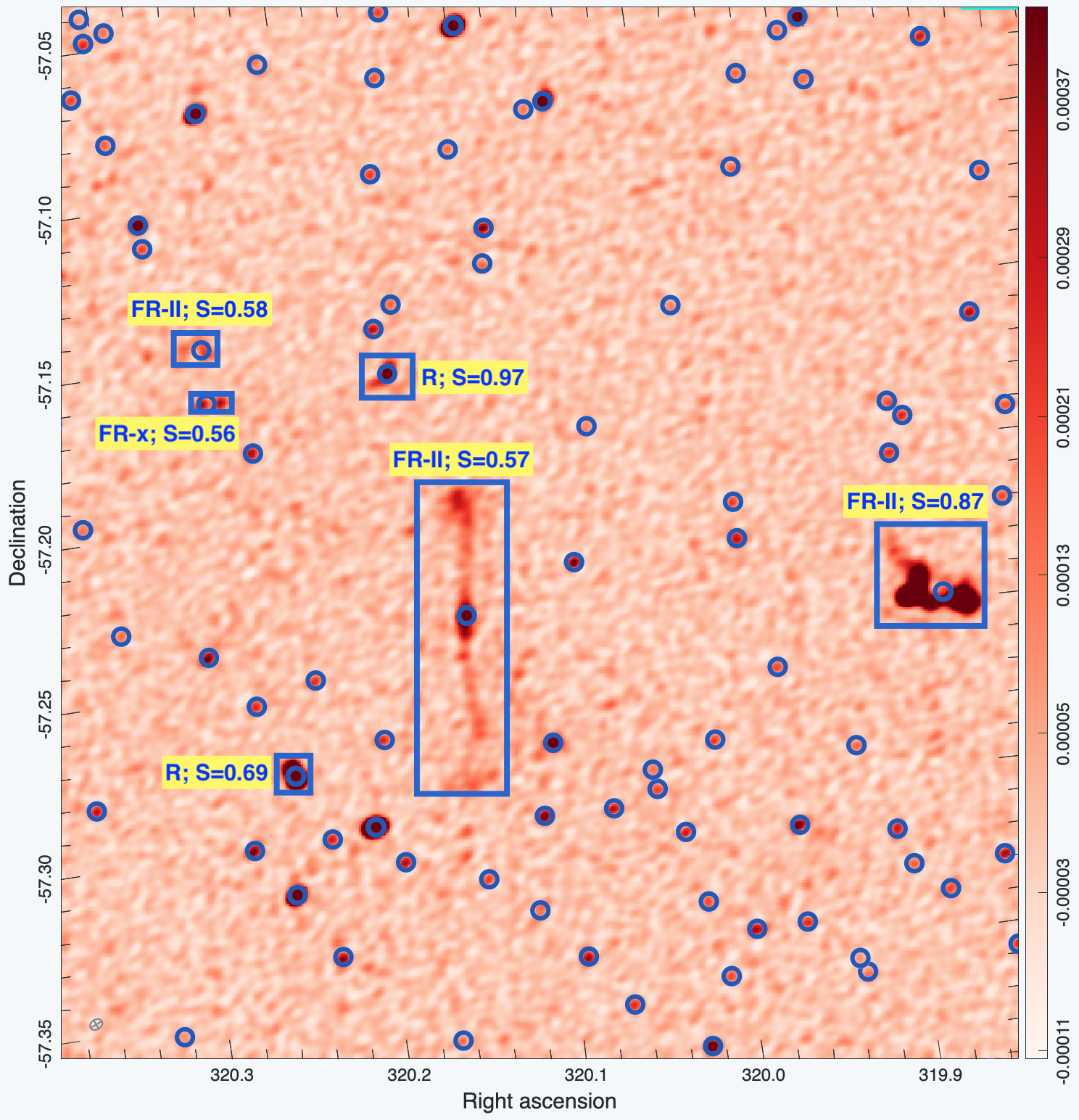}
  \caption{An example of radio galaxies from the RG-CAT catalogue, superimposed on the EMU-PS1 radio image, highlighting their host positions, bounding boxes, and classifications generated by the Gal-DINO network. See more examples in \cite{gupta24b}.}
  \label{fig:asc}
\end{figure}

\subsection{Supervised Learning}
Supervised learning is a machine learning paradigm where algorithms learn from labeled datasets to predict outcomes or make decisions based on input data. 
To develop these algorithms, appropriate labels are required for detection and classification tasks. 
We introduced RadioGalaxyNET \citep{gupta23a}, which consisted of a labeled dataset and a suite of computer vision models designed for detecting extended radio sources and their corresponding infrared host galaxies. 
The dataset included 2,800 images, comprising both radio and infrared sky channels, and contained 4,155 instances of labeled sources. 
Radio images were sourced from the EMU-PS1. 
The identification of radio galaxies and the curation of annotations, such as class labels, bounding boxes, and segmentation masks, were performed through visual inspections \citep{norris25}. 
Additionally, infrared images from the Wide-field Infrared Survey Explorer (WISE; \citep{wright10,cutri13}) were aligned with the sky positions of the radio images, and the infrared host galaxies were visually identified in the infrared images. 
We employed object detection methods for the detection and classification of radio galaxies.
Our computer vision methods included multimodal models, such as Gal-DETR\footnote{\url{https://github.com/Nikhel1/Gal-DETR}}, Gal-Deformable DETR\footnote{\url{https://github.com/Nikhel1/Gal-Deformable-DETR}}, and Gal-DINO\footnote{\url{https://github.com/Nikhel1/Gal-DINO}}, where we introduced a mechanism for the simultaneous detection of radio galaxies and infrared hosts through bounding box and keypoint detections, respectively. 
Our findings revealed that the Gal-DINO model exhibited superior performance in detecting both radio galaxies and infrared hosts. 
In addition to these models, we conducted a comparative analysis of our radio galaxy detection results with those produced by the Gal-SIOD\footnote{\url{https://github.com/Nikhel1/Gal-SIOD}}, Gal-Faster RCNN\footnote{\url{https://github.com/Nikhel1/Gal-Faster-RCNN}}, and Gal-YOLOv8\footnote{\url{https://github.com/Nikhel1/Gal-YOLOv8}} models. 
Our results indicated that Gal-DINO and Gal-YOLOv8 yielded comparable outcomes in radio galaxy detection. 

In another work, we developed the RG-CAT pipeline \citep{gupta24b} to build radio galaxy catalogues using a two-step process: first detecting compact and extended radio galaxies and their host galaxies in radio and infrared images with computer vision networks, then using these predictions to construct a catalogue.
This pipeline led to the creation of the radio galaxy catalogue for the EMU-PS1. We built the detection pipeline based on the GalDINO network, expanding RadioGalaxyNET’s 2,800 radio galaxies. 
After visual inspections, we labeled 2,090 compact and 99 rare and peculiar radio sources, resulting in a dataset of $\sim$5,000 radio galaxies for training and evaluation. 
In evaluations, 99\% of radio galaxy predictions in the centre of the images had an Intersection over Union (IoU) greater than 0.5, and 98\% of keypoint detections had less than $3^{\prime\prime}$ separation from the ground truth. 
The final catalogue included 211,625 radio galaxies: 201,211 compact and 10,414 extended, with sub-categories.
Cross-matching with infrared (CatWISE; \cite{marocco21}) and optical catalogues (DES DR2; \cite{abbott21}, DESI Legacy Surveys DR8 redshifts; \cite{zou19}, SuperCosmos; \cite{bilicki16}) revealed that 75\% of radio galaxies had counterparts in CatWISE. Additionally, 64\%, 37\%, and 11\% were matched with DES, DESI, and SuperCosmos, respectively. 
In ongoing work \citep{gupta25bprep}, we apply the RG-CAT pipeline to construct catalogues from the EMU main survey, further improving the Gal-DINO network with rotated bounding boxes and incorporating online learning.

As Gal-DINO is trained to detect both rare and peculiar radio sources, along with other radio galaxies, we further applied the trained model to the first year of the EMU main survey to identify these sources in a supervised manner \cite{gupta25a}. 
This allowed for the efficient processing of large radio datasets, enabling the rapid discovery of rare and peculiar radio sources such as Odd Radio Circles (ORCs; \cite{norris21b}) with edge-brightened rings, Galaxies with Large-scale Ambient Radio Emissions (GLAREs), and Starburst Radio Ring Galaxies (SRRGs). 
We identified 5 new ORCs, 2 candidate ORCs, 56 GLAREs, and 18 SRRGs, offering new insights into their potential formation mechanisms. 
While such tasks are typically performed using unsupervised machine learning methods--since the target objects are unknown--here we demonstrate an efficient supervised learning approach.

\subsection{Unsupervised Learning}
Unsupervised learning is a machine learning approach where algorithms identify patterns or structures in unlabeled data without explicit guidance on the desired output. 
These methods are often applied to anomaly detection tasks. We presented an unsupervised machine learning approach to search for the rarest and most intriguing sources in EMU-PS1 \citep{gupta22}. 
We utilized self-organizing maps (SOMs; \citep{kohonen82}) and accounted for affine transformations of astronomical images to mitigate the effects of rotational and flipping variances. SOMs were applied to approximately 42,000 cutouts at the positions of extended radio sources in EMU-PS1. 
The trained model was then used to map these sources into a $10\times 10$ lattice of neurons based on their relative similarity. 
The Euclidean distance metric was employed to identify the rarest and most interesting sources in the survey, among which two new ORC candidates and several other complex radio morphologies were identified.

\section{Self-supervised Learning}
Self-supervised learning is a technique where models learn representations from unlabeled data by generating their own supervision signals. Unlike traditional unsupervised learning, which focuses on finding patterns without explicit objectives, self-supervised learning uses pretext tasks to generate meaningful supervision signals. Multimodal foundation models are pretrained using self-supervised techniques to build versatile representations across text, images, and other modalities. These models can then be fine-tuned on domain-specific datasets to adapt their general knowledge for specialized tasks.
In recent work, we investigated the use of a multimodal foundation model in radio astronomy, focusing on OpenCLIP \citep{cherti23}, an open-source pre-trained model, to classify and retrieve radio sources from the EMU survey. This study aims to improve the identification and retrieval of various radio galaxies by integrating visual and textual information into a shared embedding space. We fine-tune OpenCLIP on the RadioGalaxyNET dataset, which includes extended radio galaxies and other rare and peculiar radio sources \citep{gupta25b}. The model maps radio and infrared images to a shared latent space alongside their associated textual descriptions, enabling zero-shot classification and retrieval tasks without task-specific training.
This latent representation significantly reduces storage requirements by preserving only the embeddings containing relevant information, while also accelerating data retrieval from hours to sub-seconds. Such methods have the potential to aid in task-specific data compression. This approach led to the development of a search tool for the EMU main survey, EMUSE (Evolutionary Map of the Universe Search Engine), accessible through a web-based application\footnote{\url{https://askap-emuse.streamlit.app/}}, and it can also be used locally\footnote{follow steps at \url{https://github.com/Nikhel1/EMUSE}}.

\section{Weakly-supervised Learning}
Weakly supervised learning involves training models using incomplete, imprecise, or indirect supervision, such as noisy labels or approximate annotations, offering more flexibility than fully supervised learning while relying less on explicit labeling. 
Unlike unsupervised or self-supervised learning, which do not use labeled data, weakly supervised learning bridges this gap by leveraging limited or imperfect labels. 
In a recent study, we applied a weakly-supervised deep learning method, Gal-CAM\footnote{\url{https://github.com/Nikhel1/Gal-CAM}}, to detect and classify extended radio galaxies \citep{gupta23a}. 
The algorithm was trained on weak class-level labels to generate class activation maps (CAMs; \citep{zhou16ML}), which were further refined using an inter-pixel relations network (IRNet; \citep{ahn19ML}) to produce instance segmentation masks for the radio galaxies and their infrared host positions. 
By using only class-level information for radio galaxies, we successfully detected them in the EMU-PS1 survey images. 
Our results demonstrated that weakly-supervised deep learning could accurately predict pixel-level information, such as masks or bounding boxes for extended radio emissions and the positions of infrared hosts, even without direct pixel-level labels during training. 
We evaluated our method using mean Average Precision (mAP) at a standard intersection over union (IoU) threshold of 0.5, achieving mAP$_{50}$ scores of 67.5\% for radio masks and 76.8\% for infrared host positions.

\section{Conclusions}
\label{SEC:conclusions}
The integration of machine learning techniques has significantly advanced the analysis of data from the ongoing Evolutionary Map of the Universe (EMU) radio survey of the southern sky, allowing for more efficient detection, classification, and cataloging of radio sources. 
By leveraging supervised, unsupervised, self-supervised, and weakly-supervised learning approaches, we demonstrate that how innovative methods can enhance our understanding of complex radio galaxies and uncover new astrophysical phenomena. 
Through the development of the curated dataset RadioGalaxyNET and the application of cutting-edge object detection models such as Gal-DINO, substantial progress has been made in detecting and classifying radio galaxies. 
The advancements, enabled by the catalogue construction pipeline RG-CAT and the search engine EMUSE, have further enhanced the processing of vast radio datasets from surveys like EMU, leading to the discovery of rare and peculiar radio sources. 
Machine learning’s ability to efficiently analyse large-scale data has bridged the gap between complex datasets and meaningful astronomical insights, offering a promising future for generating expedited scientific results from next-generation radio surveys. 
As these surveys grow, and with the advent of the Square Kilometre Array (SKA\footnote{\url{https://www.skao.int/en}}) era, the continuous application and refinement of machine learning models will be essential in uncovering deeper insights into the radio universe and further enhancing the exploration of cosmic phenomena.

\section{Acknowledgements}
NG acknowledges support from CSIRO’s Machine Learning and Artificial Intelligence Future Science Impossible Without You (MLAI FSP IWY) Platform.
The scientific work discussed here uses data obtained from Inyarrimanha Ilgari Bundara / the Murchison Radio-astronomy Observatory. We acknowledge the Wajarri Yamaji People as the Traditional Owners and native title holders of the Observatory site. The Australian SKA Pathfinder is part of the Australia Telescope National Facility (\url{https://ror.org/05qajvd42}) which is managed by CSIRO. Operation of ASKAP is funded by the Australian Government with support from the National Collaborative Research Infrastructure Strategy. ASKAP uses the resources of the Pawsey Supercomputing Centre. The establishment of ASKAP, the Murchison Radio-astronomy Observatory and the Pawsey Supercomputing Centre are initiatives of the Australian Government, with support from the Government of Western Australia and the Science and Industry Endowment Fund.
This paper includes archived data obtained through the CSIRO ASKAP Science Data Archive, CASDA (\url{http://data.csiro.au}).

\bibliography{ASKAP}

\end{document}